\documentclass[twocolumn,showpacs,preprintnumbers,superscriptaddress,amsmath,floatfix,amssymb,secnumarabic]{revtex4}

\newcommand{\beqn}{\begin{eqnarray}}
\newcommand{\eeqn}{\end{eqnarray}}
\newcommand{\eq}[1]{(\ref{#1})}

\newcommand{\Tr}{ {\rm Tr} \, }
\newcommand{\tr}{ {\rm Tr} \, }

\newcommand{\sign}{ {\rm sign} \,  }

\newcommand{\fm}{\mbox{fm}}
\newcommand{\Mev}{\mbox{MeV}}
\newcommand{\Gev}{\mbox{GeV}}

\newcommand{\plq}{\mbox{pl}}
\newcommand{\rt}{\mbox{rt}}

\usepackage{amsmath}
\usepackage{amstext}
\usepackage{amsfonts}
\usepackage{amssymb}
\usepackage{subfigure}
\usepackage{color}
\usepackage{epsfig}
\usepackage{bbm}

\usepackage[bookmarks,bookmarksnumbered,linktocpage,pdfstartview=FitH]{hyperref}

\begin{document}
\sloppy
\title{Magnetic polarizabilities of  light mesons  in $SU(3)$ lattice gauge theory.}

\author{E.V. Luschevskaya}
\email{luschevskaya@itep.ru}
 \address{Institute of Theoretical and Experimental Physics, Bolshaya Cheremushkinskaya 25, Moscow, 117218, Russia}

  \author{O.E.Solovjeva}
\email{oesolovjeva@gmail.com}
\address{Institute of Theoretical and Experimental Physics, Bolshaya Cheremushkinskaya 25, Moscow, 117218, Russia}
 \address{National Research Nuclear University “MEPhI” (Moscow Engineering Physics Institute),
Kashirskoe highway 31, 115409 Moscow, Russia}

 \author{O.A. Kochetkov}
\email{oleg.kochetkov@physik.uni-r.de}
\address{Institute of Theoretical and Experimental Physics, Bolshaya Cheremushkinskaya 25, Moscow, 117218, Russia}
 \address{Institute of Theoretical and Experimental Physics, Bolshaya Cheremushkinskaya 25, Moscow, 117218, Russia\\
       Institut fur Theoretische Physik, Universitat Regensburg, D-93040 Regensburg, Germany}

 \author{O.V. Teryaev}
\email{teryaev@theor.jinr.ru}
 \address{Joint Institute for Nuclear Research, Dubna, 141980, Russia}
 \address{National Research Nuclear University “MEPhI” (Moscow Engineering Physics Institute),
Kashirskoe highway 31, 115409 Moscow, Russia}      

\begin{abstract}
 
We investigate the masses (ground state energies) of neutral pseudoscalar  and vector meson
in  $SU(3)$ lattice gauge theory in strong abelian magnetic field. The energy  of $\rho^0$ meson with zero spin projection $s_z=0$ on the axis of the external magnetic field decreases, while the energies with non-zero spins $s_z=-1$ and $+1$ increase with the field.
The energy of $\pi^0$ meson decrease as a function of the magnetic field.
 We calculated the  magnetic polarizabilities of  pseudoscalar and vector mesons for  lattice volume $18^4$.
 For $\rho^0$ with spin $|s_z|=1$  and $\pi^0$ meson the extrapolations to zero lattice spacing  have been done.   We  do not observe any evidence  in favour of tachyonic mode existence. 
 \end{abstract}

\maketitle
\section{Introduction}

 \vspace{0.2cm}

 Quantum Chromodynamics in   abelian magnetic filed of hadronic scale is a reach area for exploration.
The investigation of strongly interacting quark-hadronic matter in such field has a deep fundamental meaning.
Today it is possible to create a strong magnetic field of $15 m^2_{\pi} \sim 0.27\, \Gev^2$ ~\cite{Skokov:2009}  in   terrestrial laboratories (ALICA, RHIC, NICA, FAIR) during noncentral heavy ion collisions. Our studies have a goal to shed light on the effects that can appear in such experiments.
The  properties of fundamental particles related to their internal structure are also very important for understanding  the effects observed at the experiments.  

Let us mention  the  most famous results concerned    QCD physics in strong magnetic fields.
The charge asymmetry of  emitted charged particles \cite{Voloshin:04:1,Voloshin:08:1,Kharzeev:08:1} in non-central collisions of gold ions  at RHIC was explained by  chiral magnetic effect \cite{Kharzeev:08:1,Luschevskaya:2009,Luschevskaya:20009}.
Many phenomenological  studies have been devoted to understanding of QCD phase structure in strong magnetic fields. Chiral perturbation theory predicts decrease of transition temperature from the confinement to the deconfinement phase with increasing Abelian magnetic field ~\cite{Agasian:2008}. External magnetic fields also   strength the chiral symmetry breaking ~\cite{Gusynin:1996,Klevansky:1989,Ebert:1999,  Fraga:2008, Goyal:1999}.  
 It was shown  in the framework of  Nambu-Jona-Lasinio model the QCD vacuum becomes a superconductor in sufficiently strong magnetic field  ($B_c=m^2_{\rho}/e \simeq 10^{16}$ Tl) along the direction of the magnetic field  
  \cite{Chernodub:2010,Chernodub:2012:08,Chernodub:2012:06,Chernodub:2011,superconductivity}. 
 The transition to the superconducting phase is accompanied
by a condensation of charged  $\rho$ mesons.   The    strong magnetic fields   can also change the order of the phase transition from the confinement   to  deconfinement phase \cite{Mizher:2010,Gatto:2010,Kashiva:2011,Kanemura:1997, Klimenko:1992}.

Lattice studies   reveal  an interesting effect such as an inverse magnetic catalysis \cite{Bruckmann:2013oba}.
According to the calculations on the lattice with two types of valence quarks in QCD  the critical temperature of the transition from confinement  phase to deconfinement phase increases slightly in a strong magnetic field \cite{Massimo.DElia:2010}.  Calculations in the theory with $N_f=2+1$  \cite{Bali:2011} with dynamical quarks showed that $T_c$ decreases with increasing magnetic field. Lattice simulations with dynamical overlap fermions in two-flavor lattice QCD also showed the decrease of the critical temperature of   confinement - deconfinement transition when the field strength grows \cite{Bornyakov:2013eya}.

Numerical simulations in QCD with $N_f=2$ and  $N_f=2+1$  indicate the strongly interacting matter in  strong magnetic field  posses   paramagnetic properties in the confinement and deconfinement phases \cite{Bonatti:2013,Bonatti:2014,Bali:2013}.   Equation of state of  quark-gluon plasma   was   investigated in  \cite{Levkova:2014}.
 
Here we continue our previous work where we studied light mesons in $SU(2)$ lattice gauge theory \cite{Luschevskaya}.
We extend this analysis to the  $SU(3)$ lattice gauge theory  which is more realistic, and calculate the ground state energies of the light mesons as a function  of the magnetic field value depending on their spin.
Our previous results are in a qualitative agreement with the results of this work.
We also calculate several hadronic characteristics such as magnetic polarizabilities of light neutral pseudoscalar and vector mesons.
 The magnetic polarizability is an important physical quantity which reveals the internal structure of a particle in external magnetic field. We also made the extrapolation to zero lattice spacing where it was possible.
 Our approach is numerically expensive so we do not take into account dynamical  quarks.

 Several articles are devote to the study of meson masses in  strong magnetic field. The masses of $\rho$ mesons have  been calculated according to the relativistic quark-antiquark model in \cite{Simonov:2013}. Lattice study is given  in \cite{Hidaka:2012} in the approach with dynamical quarks and agree with our data for $eB<1\ \Gev^2$. Phenomenological study was made in \cite{Liu:2014}. For the case  of non-zero spin our results in a qualitative agreement with the results  \cite{Simonov:2013,Hidaka:2012,Liu:2014}. For zero spin our data agree with these results only for small magnetic fields.

\section{Details of calculations} 
 \label{Setup}

\vspace{0.2cm}

For generation of $SU(3)$ gauge configurations the tadpole improved Symanzik action was used

\begin{equation}
S=\beta_{imp} \sum_{\plq} S_{\plq}-\frac{\beta_{imp}}{20 u^2_0}\sum_{\rt}S_{\rt},
\label{action}
\end{equation}

where $S_{\plq,\rt}=(1/3)\Tr(1-U_{\plq,\rt})$ is the lattice plaquette (denoted as $\plq$) or 1$\times$2 rectangular loop ($\rt$), $u_0=(W_{1\times1})^{1/4}=\langle(1/3)\Tr U_{\plq}\rangle^{1/4}$   is the tadpole factor, calculated at zero temperature \cite{Bornyakov:2005}. 
This action suppresses ultraviolet dislocations which lead to the appearance of non-physical near zero-modes of Wilson-Dirac operator.

Next, we solve Dirac equation numerically
\begin{equation}
D \psi_k=i \lambda_k \psi_k, \  \ D=\gamma^{\mu} (\partial_{\mu}-iA_{\mu}) 
\label{Dirac}
\end{equation}
and find eigenfunctions $\psi_k$ and eigenvalues $\lambda_k$ for a test quark in the external gauge field $A_{\mu}$.
We find eigenmodes of Dirac operator to calculate the correlators. From the correlators we obtain ground state energies.
For the calculation of the fermion spectrum we use the Neuberger overlap operator  ~\cite{Neuberger:1997}.
This operator allows to investigate the theory in the limit of massless quarks without  chiral symmetry breaking  and can be written in this form 
\begin{equation}
D_{ov}=\frac{\rho}{a}\left( 1+D_W/\sqrt{D^{\dagger}_W D_W}  \right).
\label{overlap}
\end{equation}
 $D_W=M-\rho/a$ is the Wilson-Dirac operator with a negative $\rho/a$, $a$ 
is the lattice spacing in physical units, $M$ is the Wilson term.
Fermion fields obey periodical boundary conditions in space and antiperiodical boundary conditions in time. 
Sign function
\begin{equation}
D_W/ \sqrt{(D_W)^{\dagger} D_W}=\gamma_5 \sign(H_W),
\label{sign_function}
\end{equation}
is calculated using minmax polynomial approximation, where $H_W=\gamma_5 D_W$ is   hermitian Wilson-Dirac operator.
We investigate behaviour of the meson ground energy state in background gauge field, 
which is a sum of non-abelian $SU(3)$ gluon field and $U(1)$ abelian uniform magnetic field.
Abelian gauge fields interact only with quarks. In our calculations we have neglected the contribution of dynamical quarks. Therefore  we add  the magnetic field only in overlap Dirac operator.
For this reason we use the following ansatz:
\begin{equation}
A_{\mu \, ij}\rightarrow A_{\mu \, ij} + A_{\mu}^{B} \delta_{ij},
\label{exchange}
\end{equation}
where
\begin{equation}
 A^B_{\mu}(x)=\frac{B}{2} (x_1 \delta_{\mu,2}-x_2\delta_{\mu,1}).
\end{equation}
In order to make this substitution consistent with fermion boundary conditions, one should use the twisted boundary conditions \cite{Al-Hashimi:2009}.
Magnetic field is directed along $z$ axis and its value is quantized
\begin{equation}
qB=\frac{2\pi k}{(aL)^2}, \ \ k \in \mathbb{Z},
\label{quantization}
\end{equation}
where $q = - 1/3\, e$.

The quantization condition implies that the magnetic field has a minimal value $\sqrt{eB_{min}} = 380\ \Mev$ for $18^4$ lattice volume and $a = 0.125\ \fm$.
We are far from saturation regime, where $k/(L^2)$ is not small because we use $k$ between $0$ and $32$. 
For the inversion of overlap Dirac operator we use Gaussian source (with radius $r = 1.0$ in lattice units in space and time direction) and point receiver (the quark position smoothed with Gaussian profile).
Our simulations have been carried out on symmetrical lattices with lattice volume  $16^4$, lattice spacing $0.105\ \fm$ and lattice volume $18^4$, lattice spacings $a=0.084\ \fm,\ 0.095\ \fm,\ 0.105\ \fm,\ 0.115\ \fm$ and $0.125\ \fm$.
We use statistically independent configurations of gluonic fields for the every value of the quark mass.

\section{Observables}
 \label{Observables} 
 
\vspace{0.2cm}

We calculate the following observables in coordinate space and background gauge field $A$
\begin{equation}
\langle\psi^{\dagger}(x) O_1 \psi(x) \psi^{\dagger}(y) O_2 \psi(y)\rangle_A,
\label{observables}
\end{equation}
where  $O_1, O_2=\gamma_5,\, \gamma_{\mu}$ 
are Dirac gamma matrices, $\mu, \nu=1,..,4$ are Lorenz indices, 
$x=(\textbf{n}a, n_ta)$ and $y=(\textbf{n}^{\prime}a, n^{\prime}_t a)$ are  coordinates on the lattice.
The spatial lattice coordinate $\textbf{n},\textbf{n}^{\prime}\in \Lambda_3=\{(n_1,n_2,n_3)|n_i=0,1,...,N-1\}$ , $n_t,n^{\prime}_t$ are the numbers of   lattice sites in the time direction.
In  Euclidean space $\psi^{\dagger}=\bar{\psi}$.
In order to calculate the observables \eqref{observables} we calculate the quark propagators in coordinate space.
For the $M$ lowest eigenmodes massive Dirac propagator is represented by the following sum:
\begin{equation}
D^{-1}(x,y)=\sum_{k<M}\frac{\psi_k(x) \psi^{\dagger}_k(y)}{i \lambda_k+m}.
\label{lattice:propagator}
\end{equation}
In our calculations we use $M=50$.
For the observables \eqref{observables} the following equation is fulfilled
\begin{equation}
\langle \bar{\psi} O_1 \psi \bar{\psi} O_2 \psi \rangle_A=-\tr[O_1D^{-1}(x,y)O_2D^{-1}(y,x)]
\label{lattice:correlator}
\end{equation}
$$ +\tr[O_1D^{-1}(x,x)]\tr[O_2D^{-1}(y,y)]
$$

The first term in \eq{lattice:correlator} is the connected part, the second term is the disconnected part.
We have checked that in $SU(3)$ theory without dynamical quarks the disconnected part contribution to correlators is zero.
 We  perform Fourier transformation numerically
  \begin{equation}
 \tilde{\Phi}(\textbf{p},t)=\frac{1}{N^{3/2}}\sum_{\textbf{n}\in \Lambda_3} \Phi(\textbf{n}, n_t) e^{-ia\textbf{np}}
\label{fourie}
\end{equation}
The momenta $\textbf{p}$ has the components $p_i=2\pi k_i/(aN),\ k_i=-N/2+1,...,N/2$.
For particles with zero momentum their energy is equal to its mass $E_0=m_0$.
As we are interested in the meson ground state energy, we choose $\langle\textbf{p}\rangle=0$.
To obtain the masses we expand the correlation function to the exponential series
 $$
\tilde{C}(n_t)=\langle \psi^{\dagger}(\textbf{0},n_t) O_1 \psi(\textbf{0},n_t) \psi^{\dagger}(\textbf{0},0) O_2 \psi(\textbf{0},0)\rangle_A =
 $$
\begin{equation}
\sum_k\langle 0|O_1|k \rangle \langle k|O^{\dagger}_{2}|0 \rangle e^{-n_t a E_k},
\label{sum}
 \end{equation}
 
 \begin{equation}
\tilde{C}(n_t)= A_0 e^{-n_t a E_0} + A_1 e^{-n_t a E_1} + ... \ ,
 \label{sum2}
\end{equation}
 $A_0, A_1$ are constants, $E_0$ is the ground state energy.
 $E_1$ is the energy of first exited state, $a$ is the lattice spacing, $n_t$ is the number of  sites in the time direction.
From expansion \eq{sum2} one can see that for large $n_t$ the main contribution origins from the ground state.
Because of the periodic boundary conditions the main contribution to the ground state has the following form
$$
\tilde{C}_{fit}(n_t)=A_0 e^{-n_t a  E_0} + A_0 e^{-(N_T-n_t)  a E_0}=
$$
\begin{equation}
2A_0 e^{-N_T a E_0/2} \cosh ((\frac{N_T}{2}-n_t) a E_0).
 \label{sum33}
\end{equation}
 \begin{figure}[htb]
\begin{center}
 \includegraphics[height=8cm, angle=-90]{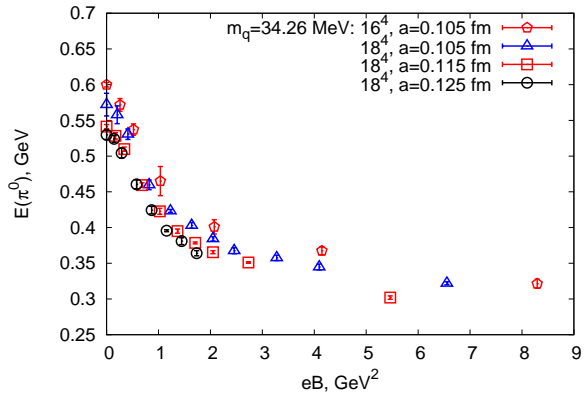}
\caption{The energy of the ground state of   pseudoscalar neutral $\pi^0$ meson obtained from the cosh. fit to the  correlator $C^{PSPS}$ as a function of the magnetic field.
The data are shown for the bare quark mass  $m_q=34.26\ \Mev$, lattice volumes  $16^4$,  lattice spacings $0.105\ \fm$ and lattice volume  $18^4$, spacings $0.105\ \fm,\ 0.115\ \fm,\ 0.125\ \fm $.}
 \label{fig:pion0}
\end{center}
\end{figure}

\begin{figure}[htb]
\begin{center}
 \includegraphics[height=8cm, angle=-90]{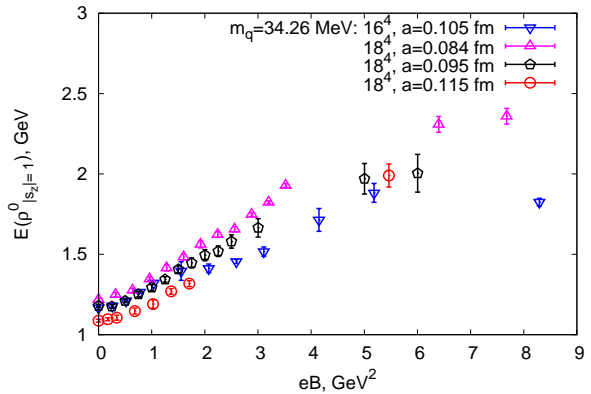}
\caption{The energy of the  $\rho^0$ ground state   obtained from the cosh. fits to the  correlators $C^{VV}_{s_z=\pm 1}$ as a function of the magnetic field for various lattice data.}
 \label{fig:rho1}
\end{center}
\end{figure}

\begin{figure}[htb]
\begin{center}
 \includegraphics[height=8cm, angle=-90]{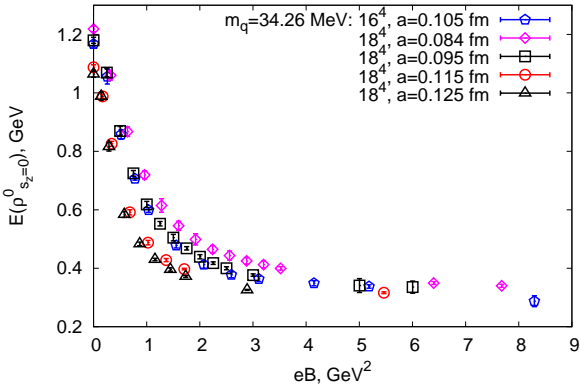}
\caption{The energy of the  $\rho^0$ ground state   obtained from the cosh. fit to the  correlator $C^{VV}_{zz}$ depending on the magnetic field value. 
The data are shown for the bare quark mass  $m_q=34.26\ \Mev$, lattice volume  $16^4$,  lattice spacing $0.105\ \fm$ and lattice volume  $18^4$, lattice spacings $0.084\ \fm,\ 0.095\ \fm,\ 0.115\ \fm,\ 0.125\ \fm $.}
 \label{fig:rho0}
\end{center}
\end{figure}

\begin{figure}[htb]
\begin{center}
 \includegraphics[height=8cm, angle=-90]{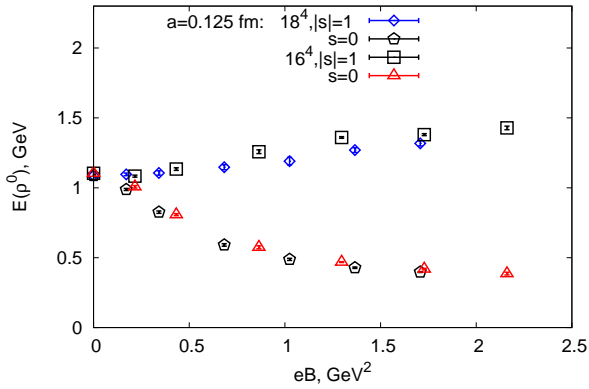}
\caption{The energy of the $\rho^0$ meson for the lattice spacing $0.125\ \fm$ and lattice volumes $16^4$ and $18^4$ for various meson spin projections.}
 \label{fig:rhovoleff}
\end{center}
\end{figure}
 
The  value of the ground state mass can be obtained by fitting the function \eq{sum33} to the lattice correlator \eq{sum}.
In order to minimize the errors and exclude the contribution of the exited states we take the values of $n_t$ from the interval $6 \leq n_t \leq N_T-6$.
Masses of the $\rho$ mesons have been obtained from correlator \eqref{observables}, where 	
$O_1,O_2=\gamma_{\mu}$. If  $O_1,O_2=\gamma_5$ we get the pseudoscalar $\pi$ 
meson. In our calculations  $u$ and $d$ quarks are degenerate.

\section{The ground state energies of mesons in strong magnetic field}

Fig.~\ref{fig:pion0} shows the ground state energy   of the neutral pion obtained from the correlator 
$C^{PSPS}=\langle \bar{\psi}(\textbf{0},n_t) \gamma_5 \psi(\textbf{0},n_t)
    \bar{\psi}(\textbf{0},0) \gamma_5 \psi(\textbf{0},0)\rangle$. On this  plot we present  the data   for the smallest  bare quark mass $m_q=34.26\ \Mev$, lattice volumes $18^4$, $16^4$ and lattice spacings $a=0.105\ \fm,\ 0.115\ \fm, \ 0.125\ \fm$. 
    The   $\pi^0$ energy decreases for the all sets of lattice data and  slightly depends on the lattice volume and lattice spacing at moderate magnetic fields.  With increase  of the field value   the lattice  effects become more strong,  the 
 wave function of   light pion becomes  of the order or exceeds the lattice size.

To obtain the energies of  neutral vector mesons with various spin projections on the axis of the external magnetic field we use the   combinations of  the correlators in various spatial dimensions. 
  \begin{equation}
C_{xx}^{VV}=\langle \bar{\psi}(\textbf{0},n_t) \gamma_1 \psi(\textbf{0},n_t)
    \bar{\psi}(\textbf{0},0) \gamma_1 \psi(\textbf{0},0)\rangle,
\end{equation}
\begin{equation}
C_{yy}^{VV}=\langle \bar{\psi}(\textbf{0},n_t) \gamma_2 \psi(\textbf{0},n_t)
    \bar{\psi}(\textbf{0},0) \gamma_2 \psi(\textbf{0},0)\rangle,
\end{equation}
  \begin{equation}
 C_{zz}^{VV}=\langle \bar{\psi}(\textbf{0},n_t) \gamma_3 \psi(\textbf{0},n_t)
    \bar{\psi}(\textbf{0},0) \gamma_3 \psi(\textbf{0},0)\rangle.
 \end{equation}
 
The mass of  $\rho^0$ meson with $s_z= 0$ spin projection is obtained from the $C_{zz}^{VV}$ correlator. 
 The combinations of correlators 
    \begin{equation}
C^{VV}(s_z=\pm 1)= C^{VV}_{xx}+C^{VV}_{yy} \pm i(C^{VV}_{xy}-C^{VV}_{yx}) 
\label{eq:CVV1}
    \end{equation}
     give the ground state energies of meson with  spin projections $s_z=+1$ and $s_z=-1$. 
     
      Fig.~\ref{fig:rho1} represents the energy of the $\rho^0$  with non-zero spin projection on the axis of the external magnetic field. The energy increases   with the field for the all sets of lattice data. At  $|eB|  \lesssim 2.5\ \Gev^2$ the energy grows quadratically and at large magnetic fields it looks like a plateau. The  terms $i C^{VV}_{xy}$ and $i C^{VV}_{yx}$ in (\ref{eq:CVV1}) are zero for the case of neutral particles so the $\rho^0$ masses with $s=-1$ and $s_z=+1$ coincide  
      which should be a  consequence of $C$-parity.
 
In  Fig.~\ref{fig:rho1} we also see  that at  $eB\gtrsim  2.5\ \Gev^2$ the data reveals more stronger dependence on the lattice spacing  than for the lower values of the field. 
 Simple estimates give the following values of magnetic fields corresponding to the lattice spacing cut-off: $2.5\ \Gev^2$ for $a=0.125\ \fm$,  $2.9\ \Gev^2$ for lattice spacing  $a=0.115\ \fm$ and 
 $eB \sim 3.5\ \Gev^2$ for $a=0.105\ \fm$ and so on. 
     
In Fig.~\ref{fig:rho0} we depict the ground state energy  which we obtained from the correlator $ C_{zz}^{VV}$  for various lattice spacings and volumes. 
 For the small magnetic fields we suggest that this energy  corresponds to the  $\rho^0$ meson with $s_z=0$. Also we find the real part of the non-diagonal terms in correlation matrix are equal to zero. At large magnetic field the mixing between $\rho^0(s_z=0)$ and $\pi^0(s=0)$ having the same quantum numbers
has to be taken into account because the branching $\rho^0\rightarrow \pi^0 \gamma$   increases with the magnetic field value.
   In this work our main goal was to calculate the polarizabilities of $\rho^0$, but not to distinguish $\rho^0$ and $\pi^0$, this can be done in the following work.
     In  Fig.\ref{fig:rhovoleff} we show for comparison the energy of $\rho^0$ meson at $a=0.125\ \fm$ for $16^4$ and $18^4$ lattices with various spins. Therefore the lattice volume effects are not large.

\begin{figure}[htb]
\begin{center}
 \includegraphics[height=8cm, angle=-90]{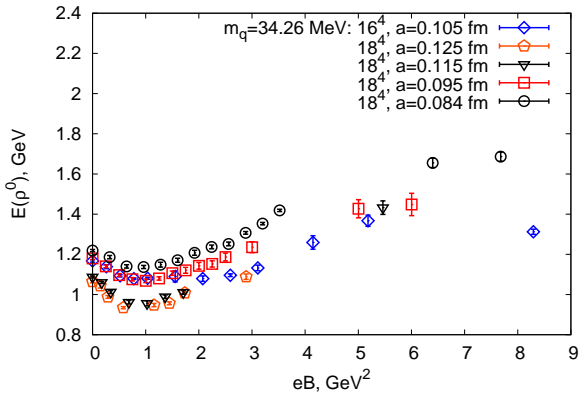}
\caption{The energy of the ground state of unpolarized vector neutral   $\rho^0$ meson  obtained from the cosh. fits to the  correlators as a function of the magnetic field.
The data are shown for lattice volume $16^4$, lattice spacing $0.105\ \fm$ and lattice volume $18^4$, lattice spacing $0.084\ \fm,\ 0.095\ \fm,\ 0.115\ \fm,\ 0.125\ \fm $ and the bare quark mass $m_q=34.26\ \Mev$.}
 \label{fig:unpolrho}
\end{center}
\end{figure}

 In Fig.\ref{fig:unpolrho} we show the $\rho^0$ energy averaged over three spin components $E(\rho^0)=(E(\rho^0_{s_z=0})+E(\rho^0_{s_z=-1})+E(\rho^0_{s_z=+1}))/3$, which corresponds to the energy of unpolarized vector meson, we suppose it has to be a constant value.
For the magnetic fields $eB<2\ \Gev^2$  the lattice spacing effects  can be a   possible reason of the   mass deviation from a constant value. 
With the diminishing of the lattice spacing the mass of unpolarized meson becomes  closer to a constant value, so it confirms the supposition that a  mixing between $\rho^0(s_z=0)$ and $\pi^0(s=0)$ states may  be weak at $eB<2\ \Gev^2$. 
 
From the assumption of constant energy of unpolarized meson and the behaviour of nonzero energy components (Fig.\ref{fig:rho1}) we can  conclude that there is no tachyonic mode for the explored range of magnetic fields, i.e. the mass of $\rho^0(s=0)$ doesn't turn to zero. We see the lattice volume effects are small and do not change this conclusion at large magnetic fields. 
The  decrease of   mass may be compensated by higher powers of $eB$ for
its values larger than $1\ \Gev$ or so. These effects can be preventing, in
particular, from mass turning to zero and possible emergence of tachyonic
mode. The same will be true also for the energies of all spin states of
 charged \cite{Luschevskaya:2014mna} $\rho$ mesons. Still, the case of neutral mesons demands further investigation and study of mixing and lattice spacing effects.

  Therefore our calculations show that there is a splitting of ground state energy of neutral vector meson in a strong abelian magnetic field that is an interesting physical effect.

 \section{Magnetic polarizabilities}

 The  polarizability of meson  is an important physical quantity for understanding of its internal  structure. The magnetic polarizability of  meson shows how current distribution responds to the external magnetic field. 
In this section we talk about the  magnetic polarizabilities of    pseudoscalar $\pi^0$ and vector  $\rho^0$  mesons.
 
 \begin{figure}[htb]
\begin{center}
 \includegraphics[height=8cm, angle=-90]{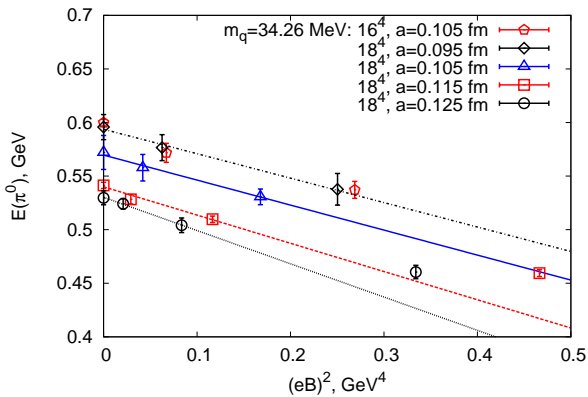}
\caption{The energy of the ground state of   pseudoscalar neutral $\pi^0$ meson   as a function of the squared magnetic field.
The data are shown for the bare quark mass  $m_q=34.26\ \Mev$, lattice volume  $16^4$, lattice spacing $0.105\ \fm$ and lattice volume  $18^4$, lattice spacings $0.095\ \fm, 0.105\ \fm,\ 0.115\ \fm,\ 0.125\ \fm $.}
 \label{fig:piB4}
\end{center}
\end{figure}

\begin{figure}[htb]
\begin{center}
 \includegraphics[height=8cm, angle=-90]{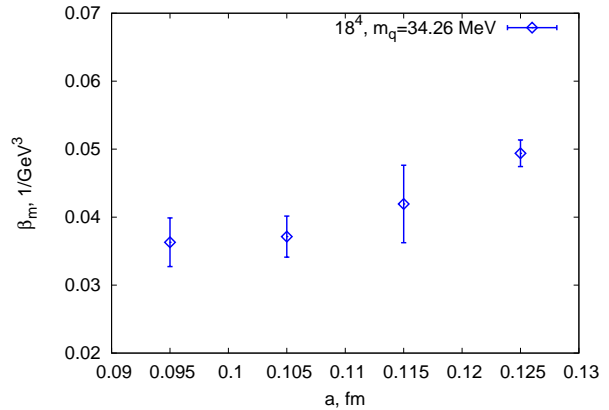}
\caption{The  magnetic polarizability of $\pi^0$ meson for various lattice spacings, the lattice volume $18^4$ and  bare quark mass $m_q=34.26\ \Mev$.}
 \label{fig:chi2fit}
\end{center}
\end{figure}

 \begin{figure}[htb]
\begin{center}
 \includegraphics[height=8cm, angle=-90]{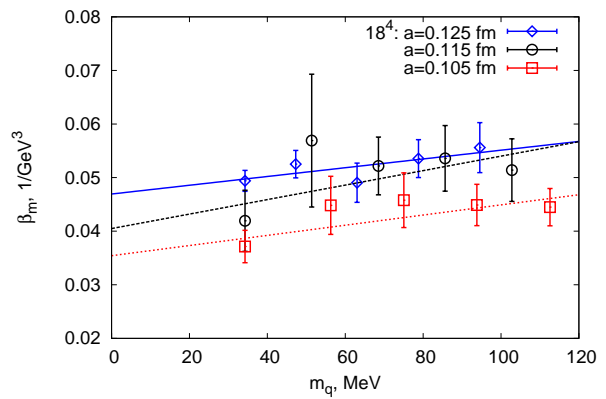}
\caption{The magnetic polarizability of pseudoscalar neutral meson $\pi^0$ versus the bare lattice quark mass for lattice volume $18^4$ and various lattice spacings.}
 \label{fig:00}
\end{center}
\end{figure}

In Fig.\ref{fig:piB4} we show the ground state energy of pion as a function of squared magnetic field $(eB)^2$ for   fields $(eB)^2<0.5\ \Gev^4$.
We fit the data  at $(eB)^2\in[0,0.3\ \Gev^4]$ by the following function
\begin{equation}
E=E(B=0)- 2 \pi \beta_m(eB)^2,
\end{equation}
where we use "natural" units $\hbar =c=1$, but $e^2=1/137$ in Gaussian units.
$E(B=0)$ and $\beta_m$ are the parameters which we find from the fit. 
We choose this interval for the fit  because we  consider   the terms $\sim (eB)^4$ give small contribution to the pion energy at  such magnetic field values.  The dashed-dotted line is for the lattice volume $18^4$, lattice spacing $a=0.095\ \fm$, the solid line corresponds to the $a=0.105\ \fm$, the dashed line is for  $a=0.115\ \fm$ and dotted one corresponds to the case of $a=0.125\ \fm$.

The obtained values of polarizabilities for various lattice spacings are summarized in  Table 1. We do not observe any functional  dependence of the magnetic polarizability on the lattice spacing, the results are  presented  in Fig.\ref{fig:chi2fit}. 
We obtain the magnetic polarizability of $\pi^0$ meson $\beta_m(\pi^0)=(0.036   \pm 0.004) \ 1/\Gev^3$ or $(2.75 \pm 0.31)\cdot 10^{-4}\ \fm^{3}$ for the   lattice spacing $0.095\ \fm$. This number agrees with sign with the result of of chiral perturbation theory  \cite{Aleksejevs:2013cda}, but its value in  $2$ times larger.  

We also explore the dependence of $\beta_m(\pi^0)$ on the bare quark mass, see Fig.\ref{fig:00}, and find the value of the magnetic polarizability slightly decreases with the diminishing  quark mass value.

\begin{table}[htpb]
\begin{center}
\begin{tabular}{|c|r|r|r|r|}
\hline
$V_{latt}$ & $a\ (fm)$            &$\beta_m^{m_q=34\ MeV}\,(\Gev^{-3})$    & Error $(\Gev^{-3})$  & $ \chi^2$/d.o.f.  \\
\hline
$18^4$     & $0.095$             &  $0.036    $      &  $0.004   $ & $ 0.0915051 $    \\
\hline
$18^4$     & $0.105$             &  $ 0.037   $       &  $0.003    $  & $0.0501177$      \\
\hline
$18^4$     & $0.115$              &  $0.042  $        & $0.006 $      & $1.03419$       \\
\hline
$18^4$     & $0.125$               &   $0.049  $     & $0.002  $    & $ 0.0130633$   \\
\hline 
\end{tabular}
\end{center}
\label{Table1}
\caption{The values of magnetic polarizability of  the   $\pi^0$ meson  
  for the bare quark mass $m_q=34.26\ \Mev$, lattice volume $18^4$ and various lattice spacings.}
\end{table}
 
\begin{figure}[htb]
\begin{center}
 \includegraphics[height=8cm, angle=-90]{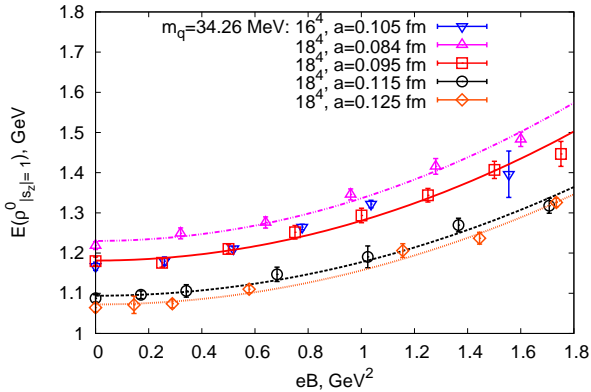}
\caption{The $\rho^0$ meson  with the fits $E=E_0(B=0)-2 \pi \beta^{|s|=1}_m (eB)^2$ to the data on the interval $eB\in[0,1.8\ \Gev^2]$. All the fits correspond to the $18^4$ lattice, the dashed-dotted  is for the lattice spacing $0.084\ \fm$, the solid line corresponds to the $a=0.095\ \fm$, the dashed one is for the  $a=0.115\ \fm$  and  dotted  line corresponds to the  case of $a=0.125\ \fm$.}
 \label{fig:rho1fit}
\end{center}
\end{figure}

\begin{figure}[htb]
\begin{center}
 \includegraphics[height=8cm, angle=-90]{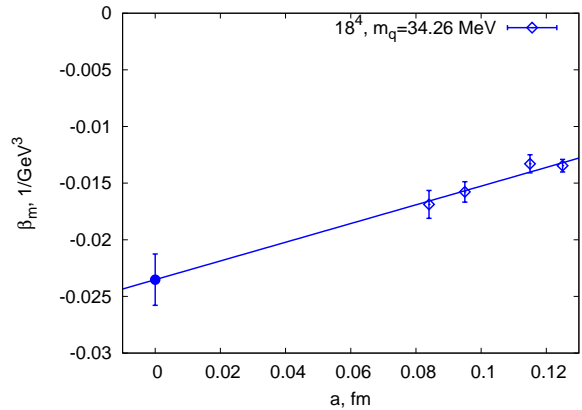}
\caption{The  magnetic polarizability of $\rho^0$ meson and its extrapolation by a linear function to zero lattice spacing for the lattice volume $18^4$ and bare quark mass $m_q=34.26\ \Mev$.}
 \label{fig:chi1fit}
\end{center}
\end{figure}

 In Fig.\ref{fig:rho1fit} in a large scale we depict the energy of the $\rho^0$  ground state with non-zero spin depending on the magnetic field for the smallest bare quark mass $m_q=34.26\ \Mev$, lattice volumes $18^4$, lattice spacings $0.105\ \fm,\ 0.115\ \fm,\ 0.125\ \fm$ and lattice volume $16^4$, spacing $0.105\ \fm$. 
To obtain the magnetic polarizability we fit our data to the function $E=E(B=0)- 2 \pi \beta_m^{|s|=1}(eB)^2$ at magnetic fields   $0\leq eB<1.8\ \Gev^2$, where the data are well described by quadratic law. $E(B=0)$ and $\beta_m^{|s|=1} $ are the unknown parameters which were found during the fitting procedure. The obtained values   $\beta_m^{|s|=1}$ for the $\rho^0$ meson with spin $|s_z|=1$ together with errors and lattice parameters  are summarized in   Table 2 and shown in Fig.~\ref{fig:chi1fit}.  

  We see the strong dependence of the results  on the lattice spacing so we make an extrapolation to the continuum limit. The extrapolation gives the value of  magnetic polarizability   $\beta_m^{|s|=1}(\rho^0)=(-0.0235 \pm 0.0023)\   1/\Gev^3$ for the  lattice volume $18^4$ and bare quark mass $m_q=34\ \Mev$. 
  The magnetic polarizability doesn't depends on the sign of the spin projection of $\rho^0$ so this quantity has a scalar nature.
  
  \begin{table}[h!tpb]
\begin{center}
\begin{tabular}{|c|r|r|r|r|}
\hline
$V_{latt}$ & $a\ (fm)$           &    $\beta_m^{m_q=34\ MeV}\,(\Gev^{-3})$ &  Error $(\Gev^{-3})$ &   $ \chi^2$/d.o.f. \\
\hline
 $18^4   $ & $0.084$             &    $-0.0169   $                &  $0.0012$  &  $1.23546$  \\
\hline
$18^4$     & $0.095$             &    $-0.0158   $                 & $0.0009$ &     $0.730133$  \\
\hline
$18^4$     & $0.115$             &    $-0.0133   $                & $0.0008$ &    $0.75398$\\
\hline
$18^4$     & $0.125$             &    $-0.0135  $                 &  $0.0006$ &   $0.831648$ \\
\hline  
$18^4$     &  $a=0$ extr.        &    $-0.0235  $                 & $0.0023$ &   $0.560989$ \\
\hline
           &                     &    $\beta^{ch.\ extr}_m\,(\Gev^{-3})$   &   &    \\
 \hline
$18^4$     & $0.115$             &    $-0.0138  $                 & $0.0005$  &  $2.64782$  \\
\hline
$18^4$     & $0.125$             &    $ -0.0161 $                  &  $0.0025  $ &  $23.8615$  \\
\hline
\end{tabular}
\end{center}
\label{Table2}
\caption{ The values of magnetic polarizability of  the   vector $\rho^0$ meson with non-zero spin
  for the bare quark mass $m_q=34.26\ \Mev$  and after chiral extrapolation, lattice volume $18^4$ and various lattice spacings.}
\end{table}

\begin{figure}[htb]
\begin{center}
 \includegraphics[height=8cm, angle=-90]{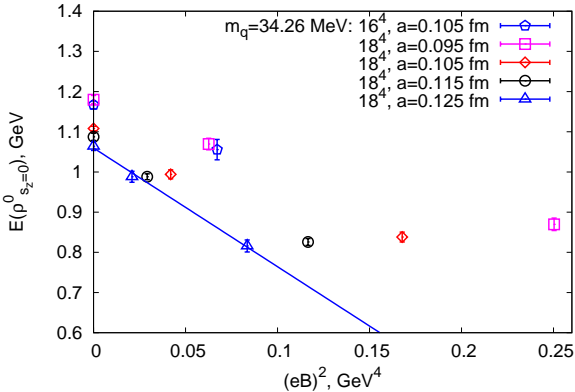}
\caption{The energy of the ground state of the vector    $\rho^0$  meson  with $s_z=0$ depending on the squared magnetic field value. The data are shown for the bare quark mass  $m_q=34.26\ \Mev$, lattice volume  $16^4$, lattice spacing $0.105\ \fm$ and lattice volume  $18^4$,lattice spacings $0.095\ \fm, 0.105\ \fm,\ 0.115\ \fm,\ 0.125\ \fm $. }
 \label{fig:rhos1B4}
\end{center}
\end{figure}

In Fig.\ref{fig:rhos1B4} the mass of $\rho^0$ meson with zero spin is depicted for the small magnetic fields. We observe the linear in $(eB)^2$ behaviour only for $(eB)^2 \in [0, 0.1\ \Gev^4]$ and get the value $\beta_m^{s=0}(\rho^0)= (0.47 \pm 0.03)\ 1/\Gev^3$ for the lattice $18^4$ and spacing $a=0.125\ \fm$. This value is   $\sim 35$ times larger and  opposite in sign compared to the   the magnetic polarizability  for non-zero spin case  $\beta_m^{|s|=1}(\rho^0)=(-0.0135 \pm 0.0006)\ 1/\Gev^3$.
Unfortunately we are limited on the magnetic field value and can't make an extrapolation to zero lattice spacing.
Large absolute value of $\beta_m^{s=0}(\rho^0)$ is most probably a lattice spacing effect. 

Positive value of the magnetic polarizability for zero spin case shows that the external magnetic field  increases  the size of  wavefunction   of the $\rho^0(s=0)$. 
 On the contrary in the case of nonzero spin the magnetic field shrinks the wavefunction of vector neutral meson, the magnetic polarizability has a negative value. 
 
The small value of the magnetic polarizability is not surprising, because  we need to apply a very large magnetic field to observe the response of the internal   structure of $\rho^0$ meson  composed of charged quarks which have a spin.

 \section{Quark mass extrapolations}

  \begin{figure}[htb]
\begin{center}
 \includegraphics[height=8cm, angle=-90]{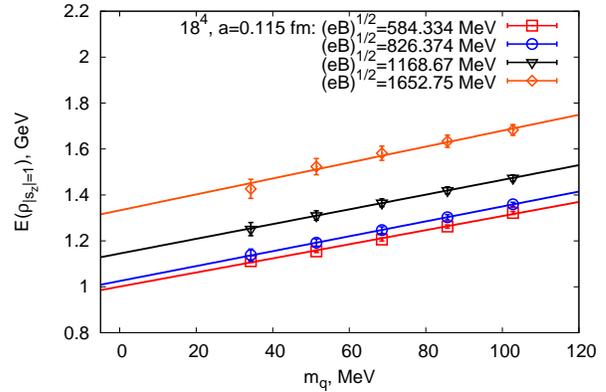}
\caption{The ground state energy of the  neutral vector $\rho$ meson spin $|s_z|=1$  for lattice volume $18^4$, lattice spacing $a=0.115\ \fm$, various quark masses and several values of magnetic fields. The extrapolation  was done by the fit \eqref{fit} to the   chiral limit. }
 \label{fig:rhomassextr}
\end{center}
\end{figure}

\begin{figure}[htb]
\begin{center}
 \includegraphics[height=8cm, angle=-90]{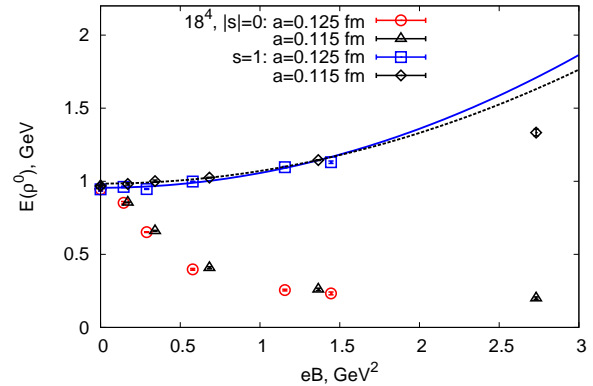}
\caption{The mass of the neutral vector $\rho^0$ meson with various spins  on
the value of external magnetic field for the lattice volume $18^4$ and lattice spacings
$a =   0.115\ \fm,\ 0.125\ \fm$ after chiral extrapolation. The solid curve is for the fit to the data at $0.125\ \fm$ lattice spacing and the dashed curve corresponds to the  data at $0.115\ \fm$. }
 \label{fig:rho1extr}
\end{center}
\end{figure}

In Fig.\ref{fig:rhomassextr} we show the quark mass extrapolation of the  neutral vector $\rho$ meson spin $|s_z|=1$  for lattice volume $18^4$ and lattice spacing $a=0.115\ \fm$. The mass of $\rho^0$ meson was calculated for several $m_q$ values in the interval $m_q a \in [0.02,0.06]$. Then we perform a fit 
by a linear function 
\begin{equation}
m_{\rho}=a_0+a_1m_q
\label{fit}
\end{equation}
and find the coefficient $a_0$ and $a_1$ from the fit and its errors   by $\chi^2$ method. Then we extrapolate $m_{\rho}(m_q)$ to  the chiral limit $m_q=0$.  
The result of such extrapolation is presented in Fig.\ref{fig:rho1extr} for the lattice volume $18^4$ and lattice spacings   $a=0.115\ \fm,\ 0.125\ \fm$.

    The masses of $\rho^0$  with zero spin smoothly decrease with magnetic field while the energies with masses spin increase with the field value.
      In Table 1 in the last three raws we also represent the values of magnetic polarizability obtained from the fits to the data after chiral extrapolation. 
The values of $\beta_m^{|s|=1}$   after quark mass extrapolations  
  agrees  with the values of   $\beta_m^{|s|=1}$  obtained for the bare quark mass $m_q=34\ \Mev$  within the errors for $a=0.115\ \fm$. For $a=0.125\ \fm$ the agreement is not so good, because of the absence of data   for the extrapolation.

\section{Conclusions}

In this work we are perform lattice QCD simulations to explore the ground state energies of   $\pi^0$ and $\rho^0$ mesons.
The mass of pseudoscalar meson diminishes with the field value.
 We observe that the energies of the ground state of neutral vector $\rho$ meson with zero spin projection on the axis of the external magnetic field decrease while the energies with non-zero spin increase as a function of magnetic field. The magnetic polarizability of $\rho^0$ meson with $s_z=0$ differs from the magnetic polarizability of $\rho^0$ meson with $|s_z|=1$.
   We consider this phenomena to be the result of the anisotropy created by the  strong magnetic field.
 The energies of $\rho^0$ with spin $s_z=+1$ and $s_z=-1$ coincides which is the  consequence  of C-parity.
 
For vector meson the magnetic polarizability   $\beta_m^{|s|=1}(\rho^0)=(-0.0235 \pm 0.0023) \ 1/\Gev^3  $   after  extrapolation to   zero lattice spacing $a=0$. For zero spin $\beta_m^{s=0}(\rho^0)$ is opposite in sign to nonzero spin case. 
  The magnetic polarizability of $\pi^0$ meson $\beta_m(\pi^0)=(0.036    \pm 0.004)\ 1/\Gev^3$ for the   lattice spacing $0.095\ \fm$.
We consider   mixing between   $\pi^0$ and $\rho^0(s=0)$ states  not strong at $eB<2\ \Gev^2$, this is the subject for the further detailed investigation. 
 We   do not observe any evidence  in favour of tachyonic mode existence.

\section{Acknowledgments} 
 
  This work was carried out   with the financial support of Grant of President  MK-6264.2014.2, FRRC grant of Rosatom SAEC  and Helmholtz Assotiation and RFBR grant 14-02-00395 A.
  The authors are grateful to  FAIR-ITEP supercomputer center where these numerical calculations were performed. 
 O.T. is supported by   RFBR grants 12-02-00613,  14-01-00647 and in part by Heisenberg-Landau  program.
 O.K. is supported by the RFBR grant 13-02-01387-a.
 
 The authors are grateful  to  Yu. A. Simonov, V. G. Bornyakov, V. V. Braguta, M. N. Chernodub and   Pavel V. Buividovich for    discussions.

\end{document}